\documentclass{article}

\usepackage{latexsym,amsfonts,amssymb,proof}
\usepackage{fancyvrb}
\usepackage{txfonts}
\usepackage{epsf}

\newcommand{\ket}[1]{|{#1}\rangle}
\newcommand{\av}{``}

\newtheorem{definition}{Definition}[section]

\newtheorem{lemma}[definition]{Lemma}

\newtheorem{proposition}[definition]{Proposition}

\title{CHARACTERIZATION\\
OF QUANTUM STATES\\
IN PREDICATIVE LOGIC}
\author{Giulia Battilotti\\
Dept. of Pure and Applied Mathematics\\
University of Padova - Italy\\
giulia@math.unipd.it}
\date{}

\begin{document}
\maketitle
\begin{abstract}
We develop a characterization of quantum states by means of first order variables and random variables, 
within a predicative logic with equality, in the framework of basic logic and its definitory equations.
 
We introduce the notion of random first order domain and find a characterization of pure states in predicative logic and mixed states in propositional logic, due to a focusing condition. We discuss the role of first order variables and the related contextuality, in terms of sequents.
\end{abstract}

\section{Introduction}
In recent years, research in quantum computation has induced logicians to stress the role of quantum states in quantum logical models (cf. \cite{DCG}). In particular, this point of view should enlight the role of quantum superposition and entanglement in quantum information. We have recently
proposed an interpretation of quantum parallelism by sequents \cite{Ba}, which describes quantum superposition and entanglement by means of first order quantifiers
in the framework of basic logic \cite{SBF}, \cite{MS}. Such an interpretation aims to justify the quantum computational processes in logical terms, namely as a process of assertions, represented by logical sequents. 

In the present paper, in order to better focus our interpretation, we perform an analysis
of the role of first order variables in the representation of quantum states. To this aim, we introduce the notion of random first order domain and find a characterization in terms of a focusing condition, which allows the distinction between pure and mixed states. We see that assertions of quantum mechanics are  predicative whereas assertions of statistical mechanics are  propositional; in addition,  we show a correspondence with the representation of states as vectors in Hilbert spaces and as density operators. Related to variables, we discuss the role of contextuality in the representation of quantum states. Our treatment of contextuality is basically the treatment of contexts in sequent calculi. We make the hypothesis that there is a role of the variable due to
the randomness of quantum mechanics, that is not considered in usual logical judgements, and can be made explicit in presence of contextuality.
In this setting, we introduce an interpretation of the uncertainty and briefly discuss the interpretation of the entanglement link introduced in \cite{Ba}.

Our work provides an alternative interpretation of logical constants and imports the notion of random variable in sequent calculus. We aim to develop the logical system so derivable, and to investigate furtherly on the focusing condition (that, in the infinitary case, would lead to G\"odel's $\omega$-rule for arithmetic) and on the related meaning of substitution, representing here the collapse of the wave function.  

We hope that our approach can contribute, from a logical point of view, to the discussion of foundational problems of quantum mechanics \cite{Ja}, such as the meaning of its randomness. 
Our approach via  first order variables, moreover, permits to face the problem of \av variables and objectivity of the state" in quantum mechanics, even if from a perspective very far from the traditional hidden variables programs. 

In the paper, logical derivations in an informal way, that correspond to the direct use of the definitory equations discussed in our model; anyway one could immediately transcribe them as formal logical derivations. Even though we think that most of our ideas could be extended to the infinitary case, we confine our attention to finite sets and discrete observables.

\section{Preliminary remarks on basic logic}

The idea that has leaded our research is to describe the information given by a physical system, physical truth, in terms of logical assertions.
An assertion, under certain assumptions, can be represented by a sequent. In particular, a temptative interpretation of the sequent
$$
\Gamma\vdash A_1,\dots, A_n
$$
is that it represents the items of information $A_1,\dots A_n$ one can achieve from a physical system, at the same time, under certain assumptions described
in the list of premises $\Gamma$ of the sequent itself. We read $\vdash$ as {\em yield}: $\Gamma$ {\em yield} info $A_1,\dots ,A_n$. Such an interpretation is taken from basic logic, that is a platform to study sequent calculi, including calculi from quantum logics, introduced in recent years (\cite{SBF}, \cite{BF}). 
In the view of basic logic, one introduces logical constants by means of {\em definitory equations} on sequents. The definitory equations describe
the translations of some metalinguistic links between assertions into logical constants of the object language.

In our model, definitory equations are used to convert the links among the information supplied by a physical system into logical constants.
The basic point for our model, that we shall see in detais in the next section, is that a metalinguistic link which can be described by the word \av{\em forall}" can be read in the description
of the state of a quantum system. 

We remind the definitory equation converting the metalinguistic link {\em forall} into the universal quantifier, introduced 
and first discussed in \cite{MS}. Let us consider any first order domain $D$ and a first order variable $z$ for its elements. One considers the
family of assertions $\Gamma\vdash A(z)$, where the premises $\Gamma$ do not depend on $z$ free. It is 
\medbreak
\centerline{
{\em forall} $z\in D$, $\Gamma\vdash A(z)$
}
\medbreak
\noindent Then the predicate $z\in D$ is considered, equivalently, a further premise besides $\Gamma$, and so {\em forall} $z\in D$, $\Gamma\vdash A(z)$ is written as the unique sequent 
$$
\Gamma, z\in D\vdash A(z)
$$
This writing is consistent with the intuitionistic interpretation of the first order quantifier (see e.g. \cite{ML} in intuitionistic type theory). Maietti\footnote{Private communication} discusses the equivalence between the two in terms of \av possibility
of substitution" of the free variable by a closed term denoting an element of the domain. As we shall see, in our model a substitution represents
a measurement, hence assuming such an equivalence means assuming the measurability in the physical system. This seems a reasonable assumption in our case, even if it has some
restriction in quantum physics, that we shall consider in the final section.  

So the definitory equation of $\forall$ we adopt is the following:
$$
\Gamma\vdash (\forall x\in D)A(x) \quad \mbox {iff} \quad \Gamma, z\in D\vdash A(z) 
$$
where $z$ is not free in $\Gamma$. In this form, one can derive the intuitionistic rules for $\forall$.

The condition on $\Gamma$ could be read as \av the additive character"\footnote{We refer to the distinction between additive and multiplicative connectives introduced in Girard's linear logic and adopted in basic logic too.} of the quantifier, that seems intrinsic to its definition.
We shall discuss in the next sections a characterization of $\forall$ with respect to the additive propositional conjunction $\&$, in terms of quantum states.
We remind here the definitory equation of $\&$, that converts the link {\em and} between two sequents with equal premise:
$$
\Gamma\vdash A\& B \qquad \mbox{iff} \qquad \Gamma\vdash A \;\mbox{{\em and}}\; \Gamma\vdash B
$$

Moreover, we adopt the multiplicative disjunction, here denoted by the multiplicative symbol $\ast$, to represent the contemporary presence of two items of independent information following from $\Gamma$, consistently with our interpretation of the notion of sequent. This is another kind of {\em and}, represented by the comma in the sequent. The definitory equation of $\ast$ we assume is the
following:
$$
\Gamma\vdash A\ast B,\Delta \qquad \mbox{iff} \qquad \Gamma\vdash A, B, \Delta
$$
The presence of a right context $\Delta$ at the right, in  our model, is due to the non contextual character of the treatment of independent information. We specify a non empty context in the definition of the multiplicative constant $\perp$:
$$
\Gamma\vdash A_1,\dots ,A_n, \perp \qquad \mbox{iff} \qquad \Gamma\vdash A_1, \dots , A_n
$$
adopted to represent the uncertainty.

Context-sensitiveness is proper of a different treatment of information with entanglement. We briefly discuss this point in the last section, where we give
a predicative definitory equation which can extend the action of the quantifier, in a paraconsistent setting. In this case, the language for our assertions is forced to go beyond the language of sequents.

So far we have defined connectives at the right side of the sequent. In basic logic, one defines the {\em dual} connectives symmetrically, at the left side of
the sequent (see \cite{SBF}). A discussion on the role of symmetry and duality in the representation of quantum states is developed in a forthcoming work. Here we remind the definitory equations of the additive disjunction $\vee$ and of the existential quantifier $\exists$, in the form adopted in the paper:
$$
\Gamma, A\vee B\vdash \Delta \quad \mbox{iff} \quad \Gamma, A\vdash \Delta \quad \Gamma, B\vdash \Delta
$$ 
$$
\Gamma, (\exists x\in D)A(x)\vdash \Delta \quad \mbox{iff} \quad \Gamma, A(z), z\in D\vdash \Delta 
$$
Finally, we shall adopt the following Leibnitz-style definitory equation of the equality predicate, introduced in the framework of basic logic by  Maietti (see \cite{Ma}):
$$
\Gamma', \Gamma(t/s), s=t \vdash \Delta(t/s), \Delta' \quad \mbox{iff} \quad\Gamma', \Gamma \vdash \Delta, \Delta'
$$
 
\section{Logical assertions describing physical states}
Since we aim to represent the information contained in a physical system,  we need to refer to the measurement of the values of the observables
in a certain state  of the system itself.

Let us consider any physical system $\cal A$ and an observable $\cal O$. In order to know the state of the system, we need to measure the value of the observable, under certain measurement assumptions. Let us assume first that the measurement assumptions determine the measurement outcome. We now see that the representation
of the information by means of a sequent is very direct in this case, since determinism is represented very well by the relation $\vdash$ ({\em yield}).
The fact that $\cal A$ is found in state $s$ corresponds to the assertion: 

\noindent \av the measurement 
assumptions and the value of the outcome $v$ yield that $\cal A$ is in state $s=s(v)$". 

\noindent We summarize the proposition \av the outcome of a measurement of $\cal O$ on $\cal A$ has the value $v$" by  $O(v)$, and the proposition \av$\cal A$ is in state $s(v)$" by $A(v)$, $v$ being a term of the language. We summarize all the measurement assumptions into the list of propositions $\Gamma$. 
Then, the propositional formula $O$ attributes a value to $\cal O$, the propositional formula $A$ attributes a value to $\cal A$ and our assertion on the state of $\cal A$ is converted formally into the sequent $\Gamma, O(v)\vdash A(v)$.

Before measurement, the value of the observable is unknown. We represent it by a free variable $x$, so our assertion concerning the state of the system has the form $\Gamma, O(x)\vdash A(x)$. After measurement, the value of the observable is represented by a closed term $t$, the closed predicates $O(t)$ and $A(t)$ attribute a value to $O$ and a state to $\cal A$. Our assertion on the state of the system is converted into 
\begin{equation}
\label{cl}
\Gamma, O(t)\vdash A(t)
\end{equation} 
In logic, such a conversion is performed by substituting the variable  by the term in the sequent $\Gamma, O(x)\vdash A(x)$. This is permitted by the substitution rule:
$$
\infer [subst]{\Gamma, O(x/t)\vdash A(x/t)}{\Gamma, O(x)\vdash A(x)}
$$
Hence a substitution describes a measurement. 
When $t$ is measured as an outcome, the proposition $O(t)$ is true. This is represented by the sequent $\vdash O(t)$. Then we can assert that the assumptions
$\Gamma$ yield that $\cal A$ is found in  state $s(t)$, that is represented by the sequent $\Gamma\vdash A(t)$. In sequent calculus, it is obtained by cutting the premise $O(t)$ in  $\Gamma, O(t)\vdash A(t)$.

We now extend the same schema to the general case, in which the measurement assumptions do not determine the measurement outcome. This enables us to deal with quantum systems too. Not surprisingly, some significant variations are required. In case of non determinism,
the information on the state of the system prior to measurement that can be achieved after a single measurement is not relevant. We need to consider a measurement
process under the same measurement assumptions.  
The outcome of a measurement process is a random variable $Z$. So to say, the instantiation of the variable describing the value of the observable gives another kind of variable, the random variable. If determinism is treated as a particular case, it is the \av constant random variable". 

To describe the information obtained from the measurement process, we avoid to consider $Z$ itself as a new term of the language, and prefer to keep ourselves in a first order language, in which we characterize a particular kind of first order domains, that we term
{\em random first order domains} (abbreviated {\em r.f.o.d.}). For each random variable $Z$, we consider  the set of its outcomes, namely the set of pairs 
$$
D_Z\equiv \{z=(s(z), p\{Z=s(z)\})\}
$$
where $s(z)$ is a state associated to an outcome (namely, a state associated to a single measurement in the measurement process\footnote{In quantum measurements, one can have more than one state associated to the value of the observable, in the degenerated case.});  $p\{Z=s(z)\}> 0$ specifies the frequency of $s(z)$ in the measurement process. The random first order domain $D_Z$ characterize the random variable of the measurement process.

So we describe the random variable $Z$ by means of the random first order domain $D_Z$ and of the first order variable $z$, which describes the generic outcome of $Z$.
The open first order predicate $z\in  D_Z$ (where $Z$ is determined by the measurement process) attributes a value to the observable. To describe the state of the system, we consider the proposition 
\av $A(z)$ \av $\cal A$ is found in state $s(z)$ with probability $p\{Z=s(z)\}$". We assume that our formal premises $\Gamma$ do not depend on the first order variable $z$, whose values are the outcomes of the random variable $Z$, since the measurement assumptions are fixed and cannot depend on its eventual outcome.  
The assertion on the state of $\cal A$ is: 

\noindent \av The measurement assumptions $\Gamma$ {\em yield} $A(z)$, {\em forall} $z\in D_Z$". 

\noindent
One writes such an assertion more formally, as a family of sequents joined by the metalinguistic link {\em forall} (see \cite{MS}):
\medbreak
\noindent \centerline{{\em forall} $z\in D_Z$, $\Gamma\vdash A(z)$} 
\medbreak
\noindent Now, one can import the premise $z\in D_Z$ into the sequent, as seen in the previous section (we have the measurability hypothesis), and has the following assertion concerning the state of the system:
\begin{equation}
\label{op}
\Gamma, z\in D_Z\vdash A(z)
\end{equation}
The assertion has an ambiguous status with respect to (\ref{cl}): on one side it is its analogous, since the value of the observable, namely the random variable, is fixed.
On the other side, it consists of open predicates. We close them by applying the 
the definitory equation of the universal quantifier:
$$
\Gamma\vdash (\forall x \in D_Z)A(x )\;\equiv\; \Gamma, z\in D_Z\vdash A(z)
$$
The quantifier $\forall $ acts as a glue which creates a new object, namely 
the proposition 
$$
(\forall x \in D_Z)A(x)
$$ 
We claim that the predicative closed formula $(\forall x \in D_Z)A(x)$ attributes a state to $\cal A$.

\subsection{Pure and mixed states}
One could immediately make the objection that a quantum state cannot be characterized by the statistical information given by a measurement.
In order to see to which extent our predicaticative formula represents the state of the given system, we now consider a substitution of
the first order variable $z$ in valid sequents of the form (\ref{op}).

Given a  system $\cal A$ and a fixed observable, for which we find r.f.o.d.  $D_Z=\{t_1\dots t_m\}$, $m\ge 1$, $t_i= (s(t_i), p\{Z=s(t_i)\})$, we describe
the state of $\cal A$ by the proposition $(\forall x\in D_Z)A(x)$. We consider the axiom of sequent calculus $(\forall x\in D_Z)A(x)\vdash (\forall x\in D_Z)A(x)$.
(read in our terms, it means that we can trivially attribute a state to our system when the measurement assumptions consist of that attribution of the state). By the definitory equation of $\forall$ (read backwards), the axiom is equivalent to the sequent:
\begin{equation}
 (\forall x \in D_Z)A(x), z\in D_Z\vdash A(z)
\end{equation}
termed reflection axiom in basic logic. Here
it is the assertion on the state of the system when the premise is the description of the state itself. 
The substitution $z/t$ yields $(\forall x \in D_Z)A(x), t\in D_Z\vdash A(t)$
from which
$$
(\forall x \in D_Z)A(x)\vdash A(t)
$$
when $t\in D_Z$ is true. The last sequent describes the transition from the information contained in $(\forall x \in D_Z)A(x)$ to the statistical information $A(t)$ obtained after measurement: one has outcome $s(t)$ with probability $p\{Z=s(t)\}$.  

\noindent The total information one can achieve from the system is described by the $m$ sequents 
$$
(\forall x \in D_Z)A(x)\vdash A(t_1)\;\dots\;(\forall x \in D_Z)A(x)\vdash A(t_m)
$$
 that are equivalent to the sequent
$$
 (\forall x \in D_Z)A(x)\vdash A(t_1)\&\dots
\& A(t_m)
$$ 
by the definitory equation of $\&$.  
Then the proposition 
$$
A(t_1)\&\dots \& A(t_m)
$$ 
represents a mixed state.

When does it represent the state of $\cal A$? In our terms, when is the proposition $(\forall x\in D_Z) A(x)$, representing the state, derivable from $A(t_1)\&\dots \& A(t_m)$?

We introduce the following definition: the domain $D_Z=\{t_1,\dots ,t_m\}$ is {\em focused with respect to the equality} $=$ in the logic we are considering, when the disjunction $z=t_1\vee \dots \vee z=t_m$ is derivable from the membership predicate $z\in D_Z$, namely  
when the sequent 
$$
z\in D_Z\vdash z=t_1\vee \dots \vee z=t_m
$$ 
is valid. 

We prove the lemma:
\begin{lemma}
Let $D_Z=\{t_1,\dots ,t_m\}$ be a focused domain. Then, for any $\Gamma$, $A$, the sequent
$$
\Gamma\vdash  (\forall x \in D_Z)A(x)
$$
is derivable from 
$$
\Gamma\vdash A(t_1)\&\dots \& A(t_m)
$$
\end{lemma}
Proof: The sequent $\Gamma\vdash A(t_1)\&\dots \& A(t_m)$ is equivalent to the $m$ sequents $\Gamma\vdash A(t_i)$ by definition of $\&$.
Then one has $\Gamma, z=t_i\vdash A(z)$,
for $i=1\dots m$, by definition of $=$ (the variable $z$ can be chosen new). 
\noindent
By the definitory equation of $\vee$, they are equivalent to the sequent
$$
\Gamma, z=t_1\vee \dots \vee z=t_m \vdash A(z)
$$
By hypothesis $z\in D_Z\vdash z=t_1\vee \dots \vee z=t_m$, and then cutting the formula $z=t_1\vee \dots \vee z=t_m$, one derives the sequent 
$$
\Gamma, z\in D_Z\vdash A(z)
$$ 
that is equivalent to $\Gamma \vdash (\forall x\in D_Z) A(x)$, by the definitory equation of  $\forall$. 
\bigbreak
\noindent 
Then a sufficient condition to answer to our question is found in the corollary: 

 \begin{proposition}
 Let $D_Z=\{t_1,\dots ,t_m\}$ be a focused domain.
 Then the sequent 
 $$
 A(t_1)\&\dots \& A(t_m) \vdash (\forall x\in D_Z) A(x)
 $$ 
 is provable for every formula $A$.
 \end{proposition}
Proof: Put $\Gamma=A(t_1)\&\dots \& A(t_m)$ in the above lemma.  
\bigbreak
It is important to stress that the interpretation of the disjunction in the focusing condition is the intuitionistic one,
namely one has $z\in D_Z\vdash z=t_1\vee \dots \vee z=t_m$ if and only if one has $z\in D_Z\vdash z=t_i$ for some $i$. This can be seen in the above proof, where the construction of the disjunction is obtained in the additive way, that gives the intuitionistic interpretation of the condition, as one could see considering the sequent calculus. This means that \av focusing is focusing which one". 

One can prove that the condition of being focused is also necessary, exploiting the duality of basic logic.
\begin{proposition}
Let us consider the domain $D=\{t_1, \dots ,t_m\}$, a language with  with equality predicate $=$, and assume that $A(t_1)\&\dots \& A(t_m)\vdash (\forall x\in D)A(x)$ holds for every $A$. Then  $D$ is focused.
\end{proposition}
Proof: Let us consider $A(x, y) \,\equiv\, x\neq y$. Then, by hypothesis, it is  $z\neq t_1 \&\dots \& z\neq t_m \vdash (\forall x\in D)z\neq x$. This is equivalent to
$z\neq t_1 \&\dots \& z\neq t_m , y\in D\vdash z\neq y$, that, by duality, gives $y\in D, z=y\vdash z=t_1\vee \dots \vee z=t_m$, from which one derives
$(\exists x\in D)z=x\vdash z=t_1\vee \dots \vee z=t_m$. Since one derives $z\in D\vdash (\exists x\in D)z=x$ (from the axiom $\vdash x=x$ by $\exists$-right rule),
one has $z\in D\vdash z=t_1\vee \dots \vee z=t_m$ cutting the existential formula. 
\bigbreak

In our interpretation, the proposition $(\forall x\in D_Z)A(x)$ represents the state of a physical system $\cal A$, the proposition $A(t_1)\&\dots \& A(t_m)$ 
represents the mixed state obtained after a non selective measurement on $\cal A$. Quantum mechanics says that, when we have a pure quantum state, the second follows from the first but they do not coincide. 
By the above propositions, the two representations proposed are equivalent if and only if the random first order domain associated to the measurement is focused.
As in the well known double-slit experiment, as soon as one tries to focus what slit the electron crosses, the interference disappears. So our representation of the state of a physical system tells us about \av its probability distribution plus its interference".

As a corollary of the last proposition, one derives also the converse of the lemma. So $D_Z$ is focused if and only if there is the equivalence between
$\Gamma\vdash A(t_1)\& \dots \& A(t_m)$ and $\Gamma \vdash (\forall x\in D_Z)A(x)$. 
This fact can be interpreted more clearly if one defines the propositional function $z\in D_Z\equiv z=t_1\vee\dots\vee z=t_m$ and then reconsiders the proof of the lemma. Its steps are then all equivalences, for they are given by the definitory equations or they are due to the definition of $D_Z$, as one can see. One direction of the equivalence says in particular that, given the set $D=\{t_1,\dots ,t_m\}$ and the $m$ judgements $\Gamma\vdash A(t_i)$, one derives $\Gamma \vdash (\forall x\in D_Z)A(x)$. We could term
such a way of getting predicative judgements from propositional judgements  {\em generalization}, since it is a way of generalizing from the data given by the experience. We have seen that judgements obtained by generalization cannot include quantum interference. In the other direction one derives the sequents $\Gamma\vdash A(t_i)$ from the assumption $\Gamma, z\in D_Z\vdash A(z)$. Such a procedure simulates a substitution. It shows that it represents something reversible and not a real collapse, in the case of focused domains. In the unfocused case, on the contrary, one needs to define
a primitive substitution rule, that is not reversible when the variable is substituted by a closed term. Summing up, we can formally state the following:
\begin{proposition}
The substitution rule applied to a variable with domain $D$ is reversible if and only if $D$ is focused.
\end{proposition}
In our terms, this is read as the non reversibility of measurements on quantum states.

We remind that a judgement of the form $\Gamma, z\in D\vdash A(z)$ can be grasped in a  way that is independent of the experience.
For example, an intuitionistic interpretation of the quantifier explains a proof of $(\forall x\in D_Z)A(x)$ in terms of a function on a first order variable, that maps the generic $z\in D$ into a proof of $A(z)$ (see \cite{ML}).
This could be the result of an abstraction from the notion $z\in D$, which allows to forget the fact that $D$ is focused and interpret it by the notion of first order variable. 
On the other side, the interpretation of the assertion $\Gamma, z\in D\vdash A(z)$ considered for our model, could suggest that there is a primitive ability in dealing with random variables rather than first order variables. This could represent a different source for intuitionistic or classical judgements too, once randomness disappears.
In the next section we shall discuss some points concerning assertions of the form (\ref{op}), but extended considering different observables or particles, and hence different variables.

\subsection{Sharp states}
\label{sm}
We consider the particular case of a state for which the random variable $Z$ is a constant, namely the outcome is $u=(s(u),1)$, and $D_Z=\{u\}$ is a singleton. Our mind is naturally led to assume the validity of the sequent 
$$
z\in \{u\}\vdash z= u
$$
For, we have an extensional concept of set, thus a singleton cannot be unfocused. In our setting, the above sequent is equivalent to
$$
A(u)\vdash (\forall x\in \{u\})A(x)
$$
This is also a quite natural assertion: we would like that a measurement with a certain result can characterize a state. 

Since no logical
rule of sequent calculi can derive such sequents, we need to assume them as axioms, if we want to agree with common sense.  In particular, they can make the 
\av wave nature" of every particle, even classical particles, evident! 
Anyway, it would be possible, at least in principle, to conceive an interpretation by sequents without assuming the axioms.

Assuming the above axioms allows  to represent selective quantum measurements too. For, if $D_Z=\{t_1,\dots ,t_m\}$, one can consider the $m$ terms $s_i=(s(t_i), 1)$ (sharp terms), namely terms which \av forget" the probability of $s(t_i)$ in the non selective measurement of the state and attribute probability $1$ to it.
Then one can consider the set $D_Z^f=\{s_1,\dots ,s_m\}$ and the propositions $A^f(s_i)$  obtained allowing a \av forgetful" substitution of the variable $z$ by $s_i$ in $A(z)$.
A forgetful substitution of $z$ by $s_i$ in the proposition $z\in D_Z$ gives: $(z\in D_Z)(z/s_i)= s_i\in D_Z^f$. The forgetful substitution so defined describes a selective quantum measurement:
$$
\infer [f-subst]{\Gamma, s_i \in D_Z^f \vdash A^f(s_i)}{\Gamma, z\in D_Z\vdash A(z)}
$$
from the conclusion of which one has $\Gamma\vdash A^f(s_i)$,
since $s_i\in D_Z^f$ is true. 

In particular, assuming the reflection axiom $(\forall x\in D_Z)A(x), z\in D_Z\vdash A(z)$ in {\em f - subst} one derives the sequent 
$$
(\forall x\in D_Z)A(x)\vdash A^f(s_i)
$$ 
which describes the collapse into the state $s(t_i)$.
From the axioms $A^f(s_i)\vdash (\forall x\in \{s_i\})A^f(x)$ one derives $(\forall x\in D_Z)A(x)\vdash (\forall x\in \{s_i\})A^f(x)$, cutting the formula
$A^f(s_i)$. Then one can measure again and re-obtain $A^f(s_i)$. This agrees with the axiomatization of quantum mechanics.

\subsection{Our representation in terms of Hilbert spaces}
An orthonormal basis $B$ of the Hilbert space of the system is associated to any measurement. Then, a random first order domain is $D_Z=\{t_1,\dots ,t_m\}$
is described by terms $t_i$ 
where $s(t_i)=\ket {b_i} \in B$. Mathematically, vectors $\ket {b_i}$ are determined only up to phase factors. For, the inner product defined in the Hilbert space does not allow to distinguish between any two orthonormal basis which differ only by phase factors.

If we identify our resulting state $s(z)$ with a vector $\ket {b_i}$ of an orthonormal basis, writing $id(s(z), \ket{b_i})$ for the identification, the
predicate $id$ can be defined in a uniform way only if we disregard phases, namely we put  $id(s(z), \ket{b_i})\;\equiv\; \ket {s(z)}\cong \ket{b_i}$,  where $\cong$ is  
the equivalence relation between vectors defined by $\ket x\cong \ket y$ iff $e^{i\phi}\ket x=\ket y$ for some phase $\phi$. 

On the contrary, if we consider the phase factors, the identification $id(s(z), \ket{b_i})$ depends on $i$ and even on $z$, then it cannot be considered as an instance of an equality predicate which allows to obtain the focusing condition on the whole random first order domain $D_Z$. In such a case, the predicative representation
$(\forall x\in D_Z)A(x)$ shows that the state is superposed, even if it does not mention its phases explicitely.

If we disregard phases, namely $D_Z$ is focused, the predicative representation $(\forall x\in D_Z)A(x)$ is equivalent to the propositional representation $A(t_1)\&\dots \& A(t_m)$, that describes the density operator, considered as a convex combination of projectors. Each formula $A(t_i)$ attributes the state $s(t_i)$ with  weight $p\{Z=s(t_i)\}$ to the system. 

Note that a pure state 
is represented by $(\forall x\in \{u\})A(x)$ in a suitable measurement basis. The axioms of the form $A(u)\vdash (\forall x\in \{u\})A(x)$ allow the identification of the state with the corresponding projector. Anyway, as we have seen above, this is not necessary from a syntactical point of view.

\section{Contextuality of quantum assertions}
In the present section we analyze some more complex assertions on physical systems, when more than one observable or more
of one particle are considered, as a kind of context-sensitive treatment of the information. 
\subsection{Uncertainty}\label{unc}
\begin{proposition}
Let us consider a system $\cal A$ and two observables $\cal O$ and $\cal O'$. Let us assume that it is possible to perform independent measurements
of the two observables. Then the assertion on the state of the system is represented by the sequent
\begin{equation}
\label{*}
\Gamma, z\in D_Z, y\in D_{Z'}\vdash A(z), A'(y)
\end{equation}
($z$ is not free in $\Gamma$ and $A'$, $y$ is not free in $\Gamma$ and $A$).
\end{proposition}
Proof: 
Assume that the result of the measurement is $D_Z=\{t_1,\dots t_m\}$ for $\cal O$  and $D_{Z'}=\{w_1\dots w_n\}$ for $\cal {O'}$. The total outcome is described
by the $mn$ assertions $\Gamma\vdash A(t_i), A'(w_j)$, $\Gamma$ the measurement assumptions. We apply the generalization procedure described in the previous section to them. 
The $mn$  assertions $\Gamma\vdash A(t_i), A'(w_j)$ are equivalent to the $\Gamma, z=t_i, y=w_j\vdash A(z), A'(y)$ $i=1\dots m$ and $j=1\dots n$, by the definitory equation of the equality predicate (one can choose $z$ and $y$ not free in $\Gamma$). By definition of $\vee$, they are equivalent to $\Gamma, \vee_i(z=t_i), y=w_j\vdash A(z), A'(y)$, $j=1\dots n$, and then to the assertion $\Gamma, \vee_i(z=t_i), \vee_j(y=w_j)\vdash A(z), A'(y)$, that is (\ref{*})
defining the propositional functions $D(z)\equiv \vee_i(z=t_i)$ and $D'(y)\equiv \vee_j(y=w_j)$ as above.
\bigbreak 

The process of generalization just described, with two different variables $z$ and $y$, is not context-sensitive, since one keeps the information $y=w_j$ as a fixed context for every $j$, when the propositional function $D(z)\equiv \vee_i(z=t_i)$ is formed. A judgement of the form (\ref{*})
preserves its non contextual origin even when treated in an abstract way in logic, which tipically performs a non contextual reasoning, witnessed by the non contextuality of sequent calculi (namely, contexts are present in the rules, so that the derivations are non contextual). 

This is not the case of quantum mechanics: when the two observables are incompatible they cannot be measured both. The uncertainty can be described as follows in terms of logical assertions. Let us consider a physical system $\cal A$ and two incompatible observables $\cal O$ and $\cal O'$. To fix  ideas, we consider a particle and the spin w.r.t. two orthogonal directions, say the $z$ and the $y$ axis. Let us consider a measurement of the spin along the $z$ axis. Then we
have an r.f.o.d. $D_Z$ and we write the following sequent which asserts the state of the system:
$$
\Gamma, z\in D_Z\vdash A(z)
$$
It cannot be extended to (\ref{*}) since we cannot measure along $z$ and $y$ at the same time. Anyway, some information can be added to it. Let us assume
we perform a selective quantum measurement on $\cal A$ (see subsection \ref{sm}). In our terms, it means to apply a forgetful substitution $f-subst$ of $z$ with $\downarrow_z$ or
$\uparrow_z$ to the sequent $\Gamma, z\in D_Z\vdash A(z)$. The final result is $\Gamma\vdash A_f(s)$ where $s$ denotes $\downarrow_z$ or $\uparrow_z$. 
Ipso facto, this leads to the total uncertainty of the information about the value
of the  spin along $y$.  Let is consider the r.f.o.d. $D_{U_Y}=\{(\uparrow_y, 1/2), (\downarrow_y, 1/2)\}$, describing the uniform distribution
of the outcomes for the $y$ axis. Let us consider the formula $\perp_Y\equiv (\forall x\in D_{U_Y})A'(x)$. It can be added to the previous information about the $z$ axis in the sequent: $\Gamma\vdash A, \perp_Y$. 

We have indicated the above formula $\perp_Y$ as a \av falsum". 
For, ideally, considering a fixed observable determines the splitting of the information one can obtain by a measurement with assumptions $\Gamma$ into two parts: the propositional formulae $A_1,\dots ,A_n$ given by a set of compatible observables (up to a maximal one), and the \av falsum" representing the quantified formulae determined by the incompatible ones. Measuring a fixed observable
and the compatible ones is like measuring them and adding the information $\perp$ which represents the uncertainty of the incompatible ones. 
This is the content of the definitory equation of the constant falsum: 
$$
\Gamma\vdash A_1,\dots ,A_n, \perp \quad \mbox{{\em iff}} \quad \Gamma\vdash A_1,\dots ,A_n
$$
Then such an equation, as we have seen for others definitory equations,  can be read in terms of the information achieved from a physical system, once it is considered relativized to a fixed observable.

\subsection{Entanglement}
The following characterization is quite immediate:
\begin{proposition}
\label{duepart}
Let us consider a compound system of two particles $\cal A$, $\cal {A'}$, and fix an observable. Then the particles are separated if and only if the assertion
on the state of the system has the same form of (\ref{*}):
\begin{equation}
\Gamma, z\in D_Z, y\in D_{Z'}\vdash A(z), A'(y)
\end{equation}
where $D_Z$ and $D_{Z'}$ are the domains relative to the measurements of $\cal A$ and $\cal {A'}$ respectively ($z$ is not free in $\Gamma$ and $A'$, $y$ is not free in $\Gamma$ and $A$).
\end{proposition}
Proof:
Let us assume that the two particles are separated. Then one can perform independent measurements on them, and have two independent random variables for the two
particles. Then one can write an assertion of the above form.

Conversely, let us assume the assertion. It is equivalent to
$$
\Gamma\vdash (\forall x\in D_Z)A(x), (\forall x\in D_{Z'})A'(x)
$$
derived applying the definitory equation of $\forall$, extended to the case with contexts at the right, that is the case of classical logic, independently to $A$ and $A'$. Then the state is attributed to $\cal A$ and $\cal {A'}$ in an independent way, consistently with the fact that the comma \av ,", in a sequent, describes a simple justaposition  of two items of information. This means that the two random variables $Z$ and $Z'$ are independent (even if they may have the same outcomes, namely $D_Z=D_{Z'}$ as sets). Then the two particles are separated. 
\bigbreak
Applying to (\ref{*}) the definitory equation of $\ast$ first and then of $\forall$,
 one derives $\Gamma\vdash (\forall x\in D_Z)(\forall x'\in D_{Z'})(A(x)\ast A'(x'))$.
 Applying the definitory equation of $\ast$ to the sequent $\Gamma\vdash (\forall x\in D_Z)A(x), (\forall x\in D_{Z'})A'(x)$, derived from (\ref{*}) as seen in the above proof, one derives
 $\Gamma\vdash (\forall x\in D_Z)A(x)\ast (\forall x\in D_{Z'}) A'(x)$.  This means that the propositions 
 $(\forall x\in D_Z)(\forall x'\in D_{Z'})(A(x)\ast A'(x'))$ and $(\forall x\in D_Z)A(x)\ast (\forall x'\in D_{Z'}) A'(x')$ both attribute a state to
 the system. Then they are equivalent. Their equivalence is distributivity of the disjunction $\ast$ w.r.t. the infinite conjunction  $\forall$ (that holds in classical logic). One can see that distributivity is provable by the non-contextuality of sequent calculus (as it has been discussed in the framework of basic logic).
Because of distributivity, the superposed state of the system, described by the proposition  $(\forall x\in D_Z)(\forall x'\in D_{Z'})(A(x)\ast A'(x'))$, is splitted
into the two propositions $(\forall x\in D_Z)A(x)$ and $(\forall x'\in D_{Z'}) A'(x')$, that are joined by a propositional connective. In our view, this represents 
a weakening of the power of quantum superposition. 

Quantum mechanics prefers a context-sensitive reasoning. In our model, we need to introduce a new link between assertions which describes the correlation between the measurement outcomes of two particles in terms of a random variable. 
Indeed, we can see what happens applying the generalization process of proposition \ref{duepart} in a particular case of entangled particles. We consider two entangled particles, represented 
in the Hilbert space $C^2\otimes C^2$, where we fix two orthogonal basis $\{\ket{v_1}, \ket{v_2}\}$ and $\{\ket{w_1}, \ket{w_2}\}$ of $C^2$, so that the state of the system is represented by a vector of the form $a_1\ket{v_1w_1}+ a_2\ket{v_2w_2}$, $a_i$  both positive reals (as is well known, this is always possible by the Schmidt decomposition). This yields that the measurement outcome is $\ket{v_i}$ for the first particle if and only if it is $\ket{w_i}$ for the second, with equal probability $a_i^2$. So, in such measurement hypothesis, we characterize a unique random variable $S$ that can be described by the random first order domain $D_S=\{i=(s(i), p\{S=s(i)\}), i=1,2\}$. The state $s(i)$ is then associated to the vector $\ket{v_i}$ for the first particle and to the vectos $\ket{w_i}$ for the second.   
In this setting, the outcomes of a measurement can be represented by the assertions: $\Gamma, \vdash A(t_i), A'(t_i)$, $i=1, 2$, where the formula $A(t_i)$ attributes state $\ket{v_i}$ to the first particle and the formula $A'(t_i)$ attributes state $\ket{w_i}$ to the second. Then , differently from proposition \ref{duepart}, the index $i$ is unique.  As in the proof of proposition \ref{duepart}, one has equivalently $\Gamma, z=t_i\vdash A(z), A'(z)$, $i=1, 2$, that is $\Gamma, z=t_1\vee z=t_2\vdash A(z), A'(z)$, namely  $\Gamma, z\in D_Z\vdash A(z), A'(z)$, defining $z\in D_Z$ as the propositional function $D(z)\equiv z=t_1\vee z=t_2$. In the assertion \,
$\Gamma, z\in D_Z\vdash A(z), A'(z)$\, 
a correlation link is not present explicitely, we need to specify it.
We label such link by \av $,_S$" (see \cite{Ba}):
$$
\Gamma, z\in D_S\vdash A(z),_S A'(z)
$$
In such form of assertion, the outcomes for the states of the two particles are equal, or one a deterministic function of the other; the 
labelled comma \av $,_S$" indicates their correlation, instead of the simple comma \av $,$", which describes the simple justaposition of two pieces of information, without correlation.

We think that studying a primitive correlation link, one could better grasp the peculiarity of quantum judgements as a form of primitive judgements based on random variables, as we have suggested above. It could represent a completion of the idea of quantum superposition. Indeed, quantum mechanics on one side can gather different states for the same particle, that is superposition, on the other can gather different particles for the same state, since it has identical particles, that in turn, as for the states, cannot be focused.

As proposed in \cite{Ba}, considering the correlation link, one can introduce a new predicative binary connective $\bowtie$, putting the definitory equation
$$
\Gamma\vdash \bowtie_{x\in D_S}(A(x);A'(x))\quad \equiv \quad  \Gamma, z\in D_S\vdash A(z),_Z A'(z)
$$
Further work  is in progress on its logical properties, the logical system so derived and on the logical characterization of Bell's states which it allows.

\bigskip
\noindent 
{\bf Aknowledgements:} to Milly Maietti for useful discussions.

\end{document}